\documentclass [manuscript,epsfig,prl,11pt,floatfix,amsmath,amssym]{revtex4}
\usepackage[english]{babel}
\usepackage{graphicx}

\usepackage[dvips]{epsfig}

\usepackage{float}

\begin{document}

\newcommand{\Xv}{{\mathbf x}}
\newcommand{\qv}{{\mathbf q}}
\newcommand{\av}{{\mathbf a}}
\newcommand{\cH}{{\cal H}}
\newcommand{\lB}{\ell_B}
\newcommand{\apa}{a_\parallel}
\newcommand{\ape}{a_\perp}
\newcommand{\Sv}{\hat{{\bf S}}}
\newcommand{\ev}{\hat{{\bf e}}}
\newcommand{\tv}{\hat{{\bf t}}}
\title{Emergent structure of multi-dislocation ground states in curved crystals}
\author{Amir Azadi}
\affiliation{Department of Physics, University of Massachusetts, Amherst, MA 01003, USA}
\author{Gregory M. Grason}
\affiliation{Department of Polymer Science and Engineering, University of Massachusetts, Amherst, MA 01003, USA}

\begin{abstract}
We study the structural features and underlying principles of multi-dislocation ground states of a crystalline spherical cap.  In the continuum limit where the ratio of crystal size to lattice spacing $W/a$ diverges, dislocations proliferate and ground states approach a characteristic sequence of structures composed of radial grain boundaries (``neutral scars"), extending radially from the boundary and terminating in the bulk.  Employing a combination of numerical simulations and asymptotic analysis of continuum elasticity theory, we prove that an energetic hierarchy gives rise to a structural hierarchy, whereby dislocation number and scar number diverge as $a/W \to 0$ while scar length and dislocation number per scar become {\it independent} of lattice spacing.  We characterize a secondary  transition occurring as scar length grows, where the $n$-fold scar symmetry is broken and ground states are characterized by polydisperse, forked-scar morphologies.

\end{abstract}

\maketitle

Understanding the ground-state order of curved, 2D crystals remains an outstanding challenge with far ranging implications, from the assembly of viral capsids~\cite{caspar, bruinsma_pnas} and multi-component lipid membranes~\cite{gompper, olvera} to the structure and stability of particle coated-droplets~\cite{bausch}. The planar, six-fold, equitriangular packing favored by isotropic interactions is incompatible with Gaussian curvature and as a consequence, topological defects are necessary features of  ground-state order in curved crystals~\cite{kleman, sadoc}.  The importance of {\it disclinations} --- points of localized 5- or 7-fold symmetry --- has long been recognized for crystals on fixed-topology surfaces, like the well-known Thomson problem~\cite{altschuler, bowick_caccuito}.  More recently, experimental~\cite{bausch, irvine}, computational~\cite{wales_09, wales_13} and theoretical~\cite{bowick, travesset, vitelli_lucks} studies have begun to recognize the importance of a related class of defects, {\it dislocations} --- ``neutral" 5-7 dipoles --- in the minimal-energy states of curved crystals, both with and without disclinations.  Unlike disclinations, the number of dislocations, $N_d$, in curved-crystal ground states grows arbitrarily large in the continuum limit --- where $W/a$ the ratio crystal size to lattice spacing diverges --- resulting in multi-dislocation chains, known as ``scars"~\cite{bausch, bowick}, that span large portions of the crystal.   While heuristic arguments have been proposed to explain the scaling of the total number of dislocations with surface curvature~\cite{bowick, irvine}, to date there is little  understanding of precisely how defects are arranged in multi-dislocation ground states and what mechanical, geometric and microscopic parameters govern these emergent structures.

\begin{figure}[b]

\includegraphics[width=0.5\textwidth]{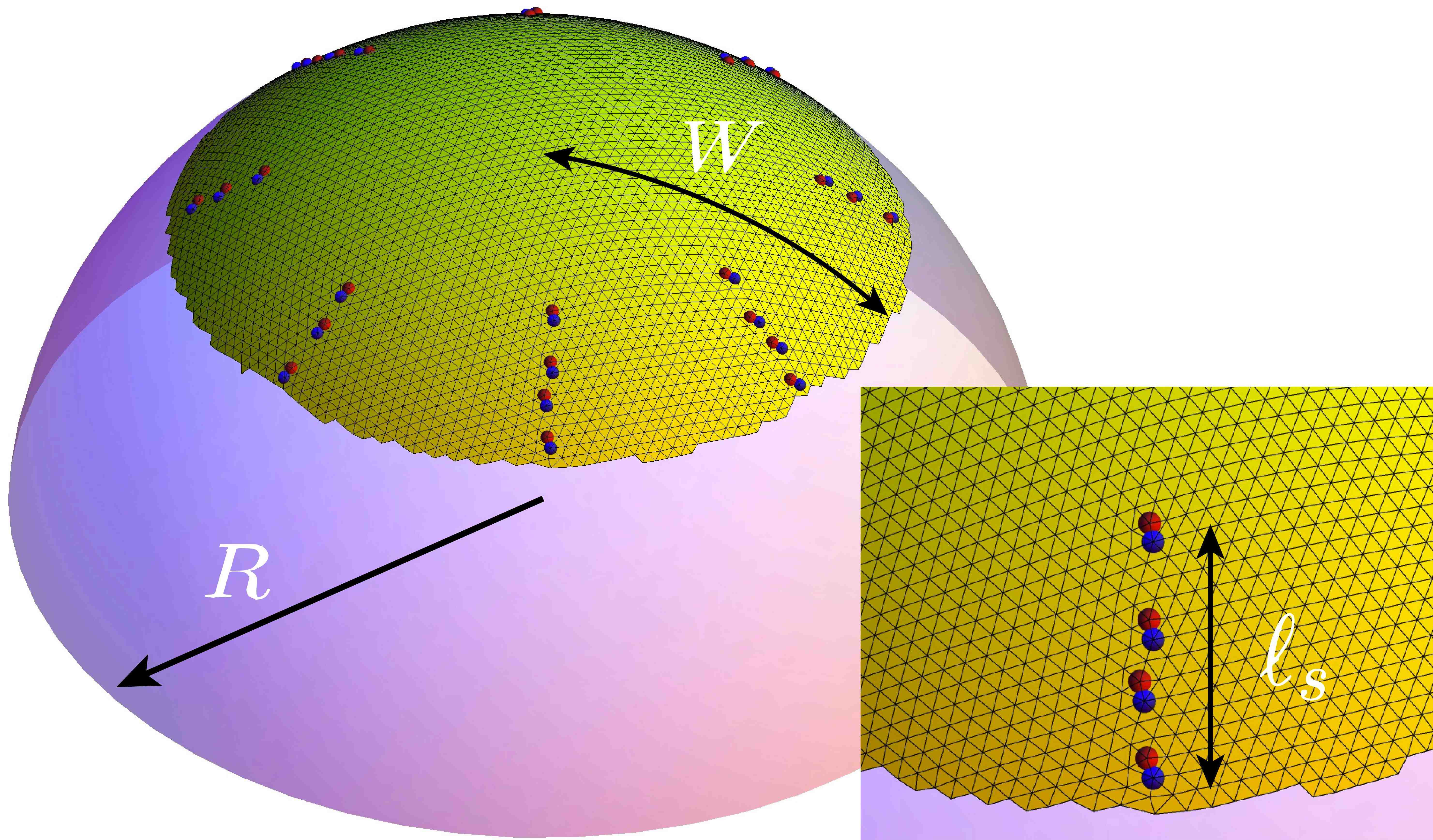}
\caption{Mesh reconstruction of an 8-scar ground state of a crystalline cap bound to sphere of radius $R$, where 5-7 dislocation ``dipoles" are shown as red and blue vertices. }
\label{cap}
\end{figure}

In this Letter, we study a continuum elasticity model of crystalline caps bound to a spherical substrate to illuminate the emergent structure of mutli-dislocation ground states in the continuum limit.  A combination of numerical and asymptotic analysis prove that as $N_d \to \infty$ the arrangement approaches a characteristic pattern:  $n_s$ radially-oriented scars extending from the crystal edge terminating in the bulk (Fig.~\ref{cap}).  An energetic hierarchy underlies the structural hierarchy characterizing these states, which was recently argued~\cite{grason_davidovitch} to parallel mechanisms of elastic pattern formation in wrinkled ultra-thin films~\cite{cerda_maha, king}, whereby certain features of the defect pattern ($N_d$ and scar length, $\ell_s$) are encoded in the mechanics of the asymptotic limit of vanishing lattice spacing, while other features (optimal scar number $n_s$) are governed by imperfect relaxation of geometric stresses by discrete dislocations.  Here, we demonstrate that optimal symmetry of $n$-fold defect patterns is selected by a competition between the distinct energetics associated with different parts of the scars, their respective {\it lengths} and {\it ends}.  Remarkably, this reveals that the asymptotic approach to the continuum limit is characterized by the divergence of {\it both} the number of dislocations {\it and} $scars$, such that $N_d/n_s$, the number of dislocations per scar, approaches a universal constant, independent of lattice spacing and defect core energy.  Finally, we present numerical evidence that the principles of this energetic hierarchy remain intact when caps are driven through a secondary structural transition which breaks the $n$-fold symmetry of the defect pattern.

We study a circular 2D crystalline ``cap" of radius $W$ bound to a rigid spherical substrate of radius $R$, subject to an adhesive, radial tension $T$ at its boundary that favors spreading of the cap over the substrate.  Our analysis is based on the continuum elasticity theory of  2D crystals, where the total energy is
\begin{eqnarray}
\label{eq: energy}
E&=&\frac{1}{2}\int dA \sigma_{ij}u_{ij}-T\Delta A .
\end{eqnarray}
For a weakly-curved crystal, elastic strain derives from in-plane displacement ${\bf u}({\bf x})$ (components in $xy$ plane) and out-of-plane defection $h({\bf x})$, with $u_{ij} = (\partial_i u_j + \partial_j u_i + \partial_i h \partial_j h)/2$, while the stress response of a hexagonal crystal is characterize by Lam\'e constants, $\lambda$ and $\mu$, $\sigma_{ij} = \lambda \delta_{ij} u_{kk} + 2 \mu u_{ij}$. The second term in (\ref{eq: energy}) represents the adhesive work where $\Delta A=W \int d\theta ~u_{r}(r=W)$ is the area change of the sheet, and $(r,\theta)$ are polar coordinates.  Dislocations are singular points, ${\bf x}_\alpha$ around which displacements increases (or decrease) by Burgers vector ${\bf b}$, corresponding to a partial row of lattice sites of width $|{\bf b}| \simeq a$ added or removed from crystal, terminating at ${\bf x}_\alpha$.  For a curved crystal possessing dislocations~\cite{nelson_book}, stress is governed by two relations, in-plane force balance, $\partial_i \sigma_{ij} = 0$, and the compatibility equation,
\begin{equation}
\label{eq: compat}
Y^{-1} \nabla^2_\perp \sigma_{ii} = - K_G - \nabla_\perp \times {\bf b} ({\bf x}) ,
\end{equation}
where $Y= 4 \mu (\lambda + \mu)/(\lambda + 2 \mu)$ is the 2D Young's modulus, $K_G=R^{-2}$ is the Gaussian curvature, and ${\bf b} ({\bf x}) = \sum_\alpha {\bf b}_\alpha \delta({\bf x} - {\bf x}_\alpha)$ is the areal Burgers density.  Note that in using eq. (\ref{eq: compat}) we assume the small-slope limit, where $|\nabla_\perp h| \approx W/R \ll 1$ and the cap covers a small (but finite) sphere fraction.  In particular, we study coverages smaller than $(W/R)_c = \sqrt{2/3} \simeq 0.82$ beyond which small-slope theory is unstable to excess 5-fold disclinations~\cite{grason_prl_10, grason_12}.

\begin{figure*}
\includegraphics[width=1.0\textwidth]{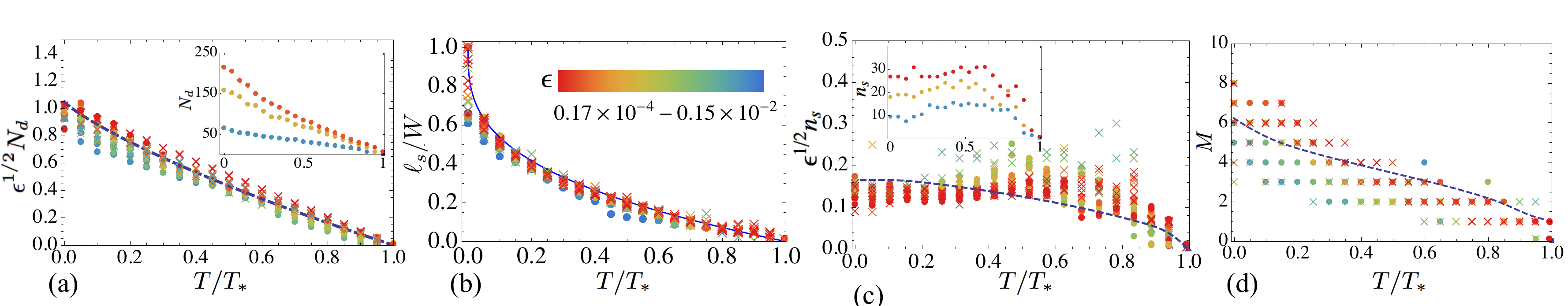}
\caption{The scaled dislocation number $\epsilon^{1/2} N_d$ (a), the length of the scarred zone $\ell_s$  (b), the scaled scar number $\epsilon^{1/2} n_s$ (c) and the number of dislocations per scar  $M$ (d) for simulated ground states of the cap are shown as functions of the reduced tension, $T/T_*$.  Insets of (a) and (c) are unscaled dislocations and scar numbers.  Results from unconstrained, ``free dislocation" and imposed $n$-fold symmetric simulations are shown respectively as crosses and filled circles.   Color scale of points in (b) correspond to dimensionless dislocaition cost $\epsilon = (b/W)^2(W/R)^{-4}$, where simulations were carried out over a range of cap sizes and curvatures:  $W/b=100-1400 $  and $W/R=0.05-0.3$.  The dashed lines indicate predictions from asymoptotic analysis of dominant and sub-dominant energetics of defect patterns.}
\label{features}
\end{figure*}

Stress in defect-free state, $\sigma_{ij}^0$, derives from geometric strains imposed by curvature and adhesive forces at the boundary, which require $\sigma_{rr} (r=W) = T$,
\begin{equation}
\label{eq: sigaxi}
\sigma^0_{rr} = \frac{Y}{16 R^2} (W^2 - r^2)+ T ; \ \sigma^{0}_{\theta \theta} =  \frac{Y}{16 R^2} (W^2 - 3r^2)+ T .
\end{equation}
Unlike the radial direction which is always tensile, in the defect-free state for sufficiently small $T$ the hoop direction becomes compressive ($\sigma_{\theta \theta}^{0}<0$) at large radii, $r>L_{0} = W/\sqrt{3} (1+2 T/T_*)^{1/2}$, where $T_* = Y/8 (W/R)^2$ is a critical tension above which the compressed zone vanishes.  Dislocations corresponding to the removal of a row extending from the defect to the boundary (i.e. ${\bf b} = b \hat{\theta}$) relax compression at the edge and lower the elastic energy, provided their cost is sufficiently low.  We characterize the susceptibility to dislocations (dubbed the ``defectivity" of the crystal~\cite{grason_davidovitch}) in terms of the ratio of dislocation self-energy, proportional to $Y b^2$, to elastic energy of the defect-free sheet, proportional to $Y W^2 (W/R)^4$,
\begin{equation}
\epsilon = (b/W)^2 (W/R)^{-4} ,
\end{equation}
which vanishes in the continuum limit $b/W \to 0$, indicating the instability of the crystal to dislocations when $T<T_*$.  We study the structure and energy of multi-dislocation configurations in this regime by superposing $\sigma^0_{ij}$ with stresses generated by multiple dislocations (${\bf b}$ aligned to hoop direction).  The self-energy of dislocations, dislocation interaction energy, and the energy associated with relaxing geometrically-induced compression derive from the free-boundary condition Greens functions of single dislocations~\cite{grason_12, azadi_grason_12} and eq. (\ref{eq: energy}) (see Supplemental Material).  For given values of tension, curvature and $b/W$, we relax the total energy by numerically adjusting defect position and number in the crystal.  For fixed $N_d$, the energy is minimized by steepest descent starting from $\sim 10^4$ random initial defect configurations.  The minimal energy multi-dislocation pattern is selected from this ensemble of ``simulated quenches".

As $T$ is reduced below $T_*$, a characteristic multi-dislocation pattern emerges:  $n_s$ evenly spaced and symmetric scars extending a distance $\ell_s$ from the edge into the cap.  For conditions shown in Fig~\ref{cap} ($W = 0.3R$, $b = 0.013W$, $T = 0.1T_*$) we find a $n_s =8$ scars of average length $\ell_s = 0.45W$, composed of $N_d = 27$ dislocations.  While optimal size and number of scars, as well as total defect number, change with both macroscopic (cap size, tension) and microscopic (Burgers vector) parameters, all simulated ground-states show spontaneous emergence of $n$-fold symmetry at the onset of scar stability, $T \lesssim T_*$.  

We now demonstrate how the features of this characteristic dislocation pattern are governed by the distribution of stress approached in the asymptotic limit $b/W \to 0$.   The ultimate stress $\sigma^{\rm d}_{ij}$  of the defect-riddled state must be significantly remodeled by dislocations from the defect free stress $\sigma_{ij}^0$, which is unstable to defects.   The stability of multi-dislocation state can be understood in terms of the Peach-Kohler force~\cite{hirth} on acting on dislocations, $f_i= b \epsilon_{ij} \sigma^{\rm d}_{j \theta}$, which implies that dislocations climbing from the boundary continue to lower the energy until defects are localized to regions where $\sigma^{\rm d}_{r \theta}= \sigma^{\rm d}_{\theta \theta} =0$.   The stable stress pattern derives from the continuum dislocation density ${\bf b}_c({\bf x}) = b \rho(r) \hat{\theta}$ that approximates defect distribution in the $N_d \to \infty$, $b\to 0$ limit, and mechanical constraints imposed by a zone of vanishing compression~\cite{grason_davidovitch}.  The axisymmetry of the areal density $\rho(r)$ implies vanishing of shear stress, while the collapse of hoop stress is governed by the solution of eq. (\ref{eq: compat}) in two radial zones:   a defect-free ($\rho=0$) {\it axisymmetric inner region} for $r<L_{\rm d}$ where the stress is identical to eq. (\ref{eq: sigaxi}) up to an overall additive constant; and an {\it outer scarred zone} ($\rho\neq 0$) for $r\geq L_{\rm d}$ where $\sigma^{\rm d}_{\theta \theta} =0$ as required by defect stability and $\sigma^{\rm d}_{rr} = T W /r$ as required by force balance and boundary conditions.   Continuity of radial and hoop components at the edge of scarred zone require an defect-free inner zone of radius
\begin{equation}
\label{eq: L}
L_{\rm d} = W-\ell_s = W(T/T_*)^{1/3} ,
\end{equation}
which predicts that scars extend {\it beyond} the original compressed zone of the defect free state since $L_{\rm d} < L_{\rm 0}$.  Like the ``far-from-threshold" analysis of wrinkling of ultra-thin elastic sheets~\cite{grason_davidovitch, king, schroll}, the asymptotic stress pattern achieved in a defect-riddled cap in the $b/W \to 0$ limit is independent of ``microscopic" features of the pattern, including $b$ and the scar number, $n_s$.  

Given this stable, compression-free pattern of stress, the dislocation distribution is determined by integrating the compatibility relation --- matching the discontinuity in $\partial_r \sigma^{\rm d}_{ii}$ at $r=L_{\rm d}$ with the dislocation density at the edge of the scarred zone --- yielding
\begin{equation}
\label{eq: rho}
\rho(r) = \frac{ \epsilon^{-1/2}}{8 W^2} \bigg[  4 \frac{r}{W}- \frac{ T}{T_*}    \Big( \frac{ W}{r} \Big)^2 \bigg] .
\end{equation}
Integrating $\rho(r)$ over the scarred zone $L_{\rm dr} \geq r \geq W$, the total dislocation number becomes,
\begin{equation}
N_d = \frac{\pi \epsilon^{-1/2}}{12 }\Big[4 (1 - T/T_*)+(T/T_*) \ln (T/T_*)\Big] .
\end{equation}
At small $T$,  $N_d \sim \epsilon^{-1/2}$  is consistent with the balance of the total edge length removed by dislocations $N_d b$ and shortening of latitudes at the outer boundary imposed by spherical geometry $\sim W (W/R)^2$, while as $T/ T_* \to 1$, boundary forces eliminate this compression, hence dislocation number vanishes in this limit $N_d \sim \epsilon^{-1/2} (T_*-T)$.  

Notably, the principle of stress-collapse in the scarred zone illustrated here is equivalent to the previously invoked notion of ``perfect screening" of Gaussian curvature by dislocations which,  for $T=0$, achieves  $\sigma_{ij} = 0$ throughout the sheet~\cite{bowick, irvine}.  Comparison to numerical simulations demonstrates that the value of the``perfect screening" distribution, and its generalization to non-zero boundary forces, is far more than heuristic, describing certain features of multi-dislocation states (length of scars and defect number) quantitatively, even for finite, but large values of $\epsilon^{-1} \sim (W/b)^2$.  In Fig.~\ref{features}a-b we compare predictions for $\ell_s$ and $N_d$ to ``free dislocation" simulations, as well as to a much larger class of numerically-optimized, fixed $n$-fold symmetry radial scar patterns, whose fewer degrees of freedom (radial positions of each dislocation ``ring") allow us to reach highly ``defective" caps, up to $\epsilon^{-1}\simeq 6 \times 10^4$ and  $N_d \approx 250$.

Unlike the dislocation number and scar length, the optimal scar number does not derive from the asymptotic stress pattern $\sigma^{\rm d}_{ij}$ in the $b/W \to 0$ limit, which is independent of $n_s$.  In \cite{grason_davidovitch}, it was shown in the limit of narrow scars ($\ell_s/W \ll 1$) that the $n_s$-degenerate energetics encoded in the elastic energy of asymptotic stress $\sigma^{\rm d}_{ij}$ correspond directly to the combination of relaxation energy per scar and the repulsive interactions between scars, which describe respectively the dominant gains and costs of multi-scar patterns.  Here, we consider {\it sub-dominant} costs of the self-energies of scars, in terms of distinct costs attributed to the {\it ends} and  {\it lengths} of scars, which describe energetics of fine-scale (intra-scar) stresses absent from the continuum limit, and more important, lift the degeneracy of the energy with $n_s$.

Scars differ from ordinary grain boundaries in that the former terminate in the bulk of crystal~\cite{bowick}.  Crossing a grain boundary implies rotation of crystal axes by $b/D$, where $D$ is the dislocation spacing.  Hence, scar ends are disclination-like singularities, points around which lattice directions rotate rapidly~\cite{hirth}, and the far-field stresses generated by scars are dominated by these end singularities.  Estimating dislocation spacing as $D = \ell_s n_s/N_d$ yields and effective disclination charge $s \approx b/D \sim (b/\ell_s)(N_d/n_s)$, and the elastic cost to introduce this charge $\ell_s \approx W$ from the cap edges becomes $\sim Y s^2 W^2$~\cite{nelson_book}.  In addition to the cost of the singular ends, grain boundary scars are characterized by a ``line tension", $\sim Y b^2/D \big[ \ln(D/b) + E_c\big]$~\cite{hirth}, where $E_c$ parameterizes the inelastic core energies of dislocations, from which we estimate
\begin{eqnarray}
\label{eq: self}
E_{\rm self} & \approx& n_s^{-1} Y(N_d b/W)^2 + Yb^2 N_d \ln\Big( \frac{N_d W}{n_s b'} \Big) \nonumber \\ &\sim &E_0 \big[ n_s^{-1} + \epsilon^{1/2} \ln (n_s \epsilon^{1/2}) \big] ,
\end{eqnarray}
where $b'$ is a renormalized core size and $E_0 \approx Y (W/R)^4 W^2$.  The elastic cost of scar tips favors a large number of low-angle scars, which is balanced by the weaker (or $\epsilon^{1/2}$) preference of line tension for dense scars (small $n_s$). This sets an optimal scar number $n_s \sim \epsilon^{-1/2} \gg 1$ that diverges in the continuum limit as $W/b \to \infty$.  As the dislocation number and scar length vary with $T/T_*$,  we expect more generally that optimal scar number of $n_s$-fold symmetric states behaves as
\begin{equation}
\label{eq: ns}
n_s = \epsilon^{-1/2} \bar{n}_s (T/T_*)
\end{equation}
where $\bar{n}_s (x)$ is dimensionless function which vanishes as $x \to 1$.  Assuming $n$-fold symmetry for all $T$, we may determine $\bar{n}_s (T/T_*)$ by numerically optimizing self-energy contributions for all $T/T_*$ (see Supplemental Material).  This prediction for optimal scar number is compared numerical ground states (both $n$-fold and ``free dislocation" simulations) in Fig.~\ref{features}c, confirming the collapse of optimal scar number to form of eq. (\ref{eq: ns}) as $\epsilon \to 0$.  Both dislocation and scar number diverge as $\epsilon^{-1/2}$, implying a universality in the approach to the continuum distribution of dislocations.  Remarkably, the number of dislocations per scar $N_d /n_s\equiv M(T/T_*)$ is predicted to approach a constant value for a given ratio $T/T_*$, {\it independent} of lattice spacing.  As shown in Fig. ~\ref{features}d, $M$ is varies weakly with tension, from $M \simeq 1$ as $T \to T_*$, to roughly 6 dislocations per scar in the absence of boundary forces ($T=0$).

\begin{figure}
\includegraphics[width=0.75\textwidth]{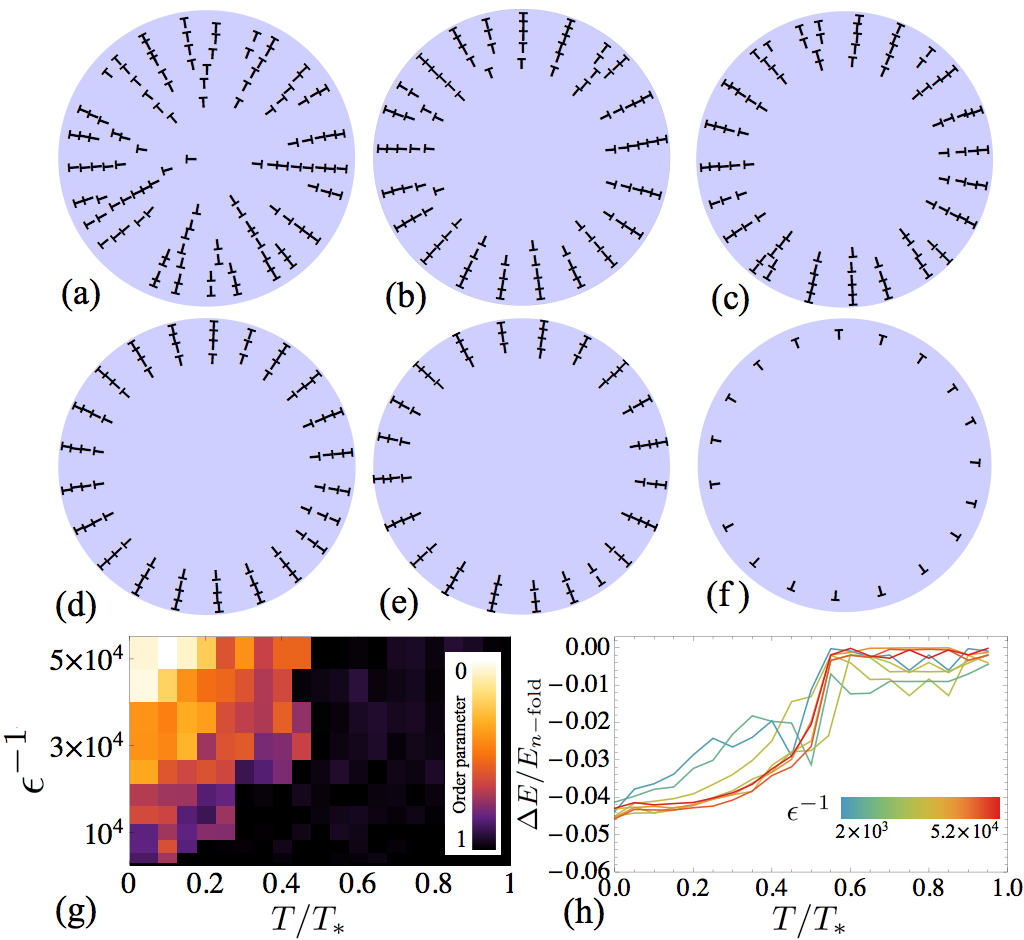}
\caption{ (a)-(f)show free dislocation ground-state configurations for $\epsilon = 0.54 \times 10^{-4}$ and sequence of increasing tension:  $T/T_* = 0, 0.1 , 0.15  , 0.3, 0.45$ and 0.7.  In (g), map of the degree of $n$-fold symmetry of dislocation pattern as measured by order parameter $S$ (defined in text), with dark and light colors showing regions of $n$-fold symmetric and polydisperse, forked-scar patterns, respectively.   In (h), relative energy difference, $\Delta E/E_{n-{\rm fold}}$, between sub-dominant energy cost of ``free dislocation" and (fixed) $n$-fold symmetric patterns normalized by sub-dominant energy as functions of reduced tension. }
\label{order}
\end{figure}  

We conclude with an analysis of the symmetry of scar patterns in our ``free dislocation" simulations (e.g. defect positions not constrained to $n$-fold patterns) examples of which are shown in the range $0\leq T < T_*$ in Fig.~\ref{order}.   We quantify the degree of $n$-fold symmetry in terms of the angular transform of simulated dislocation positions, $\bar{\rho}_m = \int dA~ e^{i m \theta} \rho({\bf x})$, and analyze the relative amplitudes of the principle non-zero mode $m = n_s$---which serves as definition of scar number of ``free dislocation" simulations---compared to higher harmonics of the distribution, $m=k n_s$.  Identical, evenly spaced scars imply $|\bar{\rho}_{n_s}|=|\bar{\rho}_{2n_s}|=|\bar{\rho}_{3n_s}|=\ldots$, and therefore, we define $S\equiv |\bar{\rho}_{2n_s}|/|\bar{\rho}_{n_s}|$ as a measure of perfect $n$-fold symmetry.  Fig. ~\ref{order}g shows the variation of $n$-fold symmetry $S$ with boundary tension and susceptibility to defects, $\epsilon^{-1}$.  Significantly, for sufficiently large tension ($T \lesssim T_*$) simulated ground states retain high-symmetry, characterized by $S \simeq 1$.  Decreasing $T$ for fixed $\epsilon^{-1}$, we find an abrupt transition to $S \ll 1$, indicating marked loss of $n$-fold symmetry, coincident with the appearance of polydisperse, or forked, scar morphologies observed for $T \to 0$ (Fig .~\ref{order}a-c).  Our simulations suggest that in the continuum limit ($\epsilon \to 0$) $n$-fold symmetric dislocation patterns become unstable to a lower symmetry, multi-scale pattern for $T \lesssim 0.4 T_*$, or equivalently, when the length of scarred zone exceeds a critical value, $\ell_s \gtrsim 0.3 W$.  

While we relegate a detailed study of this structural instability to a future publication~\cite{azadi_tobe}, we observe here that transition from $n$-fold to ``forked scar" patterns in our simulations is consistent with a transition in the {\it subdominant energetics} associated with fine-scale variations in the elastic energy.  Removing the energy encoded in the field $\sigma_{ij}^{\rm d}$ from the total energy (see Supplemental Material) Fig.~\ref{order}h compares the subdominant energies of  ``free dislocation"  to fixed $n$-fold simulations, showing the instability of $n$-fold patterns gives way to a distinct decrease in the subdominant energy by an amount ($\sim 5\%$) which saturates for large $\epsilon$.  The apparently equivalent scaling of subdominant energy with $\epsilon$ implies that the loss of $n$-fold symmetry does not alter the asymptotic, compression-free stress distribution $\sigma^{\rm d}_{ij}$ achieved in the $b/W\to 0$ continuum limit.  As a consequence, those features of the dislocation pattern determined by this asymptotic stress, the scar length and dislocation number, are not altered by the loss of $n$-fold symmetry, as we observe in Fig.~\ref{features}a-b.  Moreover, the ``scar number"  of forked-scar patterns as measured by the primary mode number of $\bar{\rho}_m$ follows the same data collapse in terms of $T/T_*$ and $\epsilon$ implied by eq. (\ref{eq: ns}) for $n$-fold symmetric patterns (Fig.~\ref{features}c-d), highlighting the more general applicability of the structural and energetic hierarchy for controlling defect patterns beyond conditions of idealized symmetry.

In summary, multi-dislocation ground states of curved crystals exhibit a characteristic sequence of patterns whose features are governed in concert, by the state of ``perfect screening" of geometrically induced stresses achievable in the singular limit $a/W \to 0$, and simultaneously, by the subdominant mechanical costs associated with the imperfect approximation of this state with a finite number of discrete defects.   Future work will reconsider long-standing questions about the asymptotic approach to continuum limit of spherical crystals at large surface coverage (e.g. the Thomson problem) which are characterize 5-fold disclinations decorated by multi-scar patterns whose optimal symmetry remains unknown.

\begin{acknowledgments}
It is a pleasure to acknowledge B. Davidovitch for essential discussions and a critical reading of this manuscript.  This work was supported by the NSF Career program under DMR Grant 09-55760 and the Alfred P. Sloan Foundation.
\end{acknowledgments}

\section{Supplementary Material}

\subsection{Effective theory of multi-dislocation caps}
Here, we provide a summary of our simulation method for continuum elastic energy of multi-dislocation patterns of crystalline caps.  We begin with the effective energy, expressed purely in terms of defect positions in the cap.  

Beginning with the continuum expression for a cap adhesively bound to a rigid sphere,
\begin{eqnarray}
E&=&\frac{1}{2}\int dA \sigma_{ij}u_{ij}-T\Delta A
\end{eqnarray}
we decompose the total stress in two components
\begin{equation}
\sigma_{ij}=\sigma^{0}_{ij}+\sigma^{D}_{ij},
\end{equation}
where $\sigma^{0}_{ij}$ represents the axisymmetric stress of the cap in the defect-free state and $\sigma^{D}_{ij}$ represents stresses generated by defects (distinct from the stress $\sigma^{{\rm d}}_{ij}$ of the ``defect riddled" ground-states defined in the main text).  The defect free stress field  $\sigma^{0}_{ij}$, eq. (3) of the main text, is the solution of the compatibility equation
\begin{equation}
\label{eq: compat}
Y^{-1} \nabla^2_\perp \sigma^{0}_{ii} = - K_G ,
\end{equation}
subject to the boundary condition $\sigma^0_{rr}(r=W)=T$.  This part of stress quantifies the cost of frustration of the confined sheet associated with axially symmetric stresses.  On the other hand the stress distribution in presence of dislocations, $\sigma^{D}_{ij}$  is governed by, 
\begin{equation}
\label{eq: compat}
Y^{-1} \nabla^2_\perp \sigma^{D}_{ii} =- \nabla_\perp \times {\bf b} ({\bf x}) =  \sum_\alpha ( {\bf b}_\alpha \times \nabla_\perp )\delta({\bf x} - {\bf x}_\alpha) , 
\end{equation} 
where ${\bf b} \times \nabla_\perp = \epsilon_{ij} b_i \partial_j$.  To maintain fixed total stress at the boundary, $\sigma^D_{ij}$ satisfies vanishing normal stress at $r=W$.  The elastic energy deriving from $\sigma^D_{ij}$ field encodes both the self-energy of dislocations and the interaction energy between dislocations.  These energies were calculated analytically in ref. \cite{azadi_grason_12} in terms of the Greens function of the biharmonic equation subject to the vanishing normal stress, where dislocations correspond to 5-7 disclination dipole.   The elastic self-energy of a single dislocation at radial position $r$ is
\begin{equation}
E^D_{self}\left(\mathbf{b},r\right)=\frac{Y ({\bf b} \cdot \hat{\theta} )^2  }{ 8 \pi^2 }\left(\frac{r}{W}\right)^{2}+\frac{Y |{\bf b}|^{2}}{8 \pi^2  }\left[\ln \Big (1-\frac{r^2}{W^2}\Big)-\ln \left(\frac{a}{W}\right)+E_{c}\right],
\end{equation}
where $E_c$ parameterizes the microscopic energy of the dislocation core (note the expression for {\it dislocation} self energy, $E^D_{self}$, should not be confused with the self-energy of {\it scars}, $E_{\rm self}$, described in the main text).  The pairwise elastic interactions between dislocations ${\bf b}_1$ and ${\bf b}_2$ at respective positions ${\bf x}_1$ and ${\bf x}_2$ take the form
\begin{multline}
E^D_{int}\left(\mathbf{b}_{1},\mathbf{x}_{1};\mathbf{b}_{2},\mathbf{x}_{2}\right)=\frac{Y}{4\pi^{2}}\bigg[-\frac{(\mathbf{b}_{1} \cdot \mathbf{b}_{2})}{2}(\ln \cos ^{2}\xi+\sin^{2}\xi) \\
 +\frac{(\mathbf{r}_{1}\times\mathbf{b}_{1})(\mathbf{r}_{2}\times\mathbf{b}_{2})}{W^2}(1-\cos^{4}\xi) 
 +\frac{(\mathbf{b}_{1}\times\mathbf{\Delta x}_{12})(\mathbf{b}_{2}\times\mathbf{\Delta x}_{21})}{|\mathbf{\Delta x}_{12}|^{2}}\sin^{4}\xi
\\ +\frac{(\mathbf{b}_{1}\times\mathbf{\Delta x}_{12})(\mathbf{b}_{2}\times\mathbf{xs}_{2})(1-r_{1}^2/W^2)+(\mathbf{b}_{2}\times\mathbf{\Delta x}_{21})(\mathbf{b}_{1}\times\mathbf{x}_{1})(1-r_{2}^2/W^2)}{(W^2-r_{1}^{2})(W^2-r_{2}^{2})+|\mathbf{\Delta x}_{12}|^{2}}\sin^{2}\xi\bigg]   \nonumber\\
\end{multline}
where $\xi$ is
\begin{eqnarray}
\cos ^{2}\xi=\frac{|\mathbf{\Delta x}_{12}|^{2}}{(W^2-r_{1}^{2})(W^2-r_{2}^{2})+|\mathbf{\Delta x}_{12}|^{2}}.
\end{eqnarray}
The coupling of the dislocation induced stresses to the curvature and tension induced stresses---cross terms $\frac{1}{2} \int dA (\sigma^D_{ij}u^0_{ij} +\sigma^0_{ij}u^D_{ij})-2 \pi W T u_r^D(W)$---lead to the ``relaxation energy" associated with release of hoop compression from the cap.  This energy is equivalently derived from  the Peach-Koehler force $f_{i}(r)=\epsilon_{ij}\sigma_{jk}(r)b_{k}$ experienced by dislocation subject to stresses $\sigma^{0}$ (and associated boundary forces).  The  relaxation of defects may be calculated from the ``climbing" of a dislocation from the edge at $r=W$ into the cap, 
\begin{equation}
E^D_{relax}(r)=b\int_{r}^{W} dr'\sigma^{0}_{\theta\theta}(r')=\frac{YW^2 b}{16 R^2}r\left[(r/W)^2-1\right]+TWb(1-r/W) .
\end{equation}
Hence the total elastic energy of the scarred crystal with $N_{d}$ dislocations on a curved surface in eq. (1) can be described by 
\begin{eqnarray}
E_{tot}=E_{0}+\sum_{\alpha=2}^{N_{d}}\sum_{\beta<\alpha}^{N_{d}} E^D_{int}\left(\mathbf{b}_{\alpha},r_{\alpha};\mathbf{b}_{\beta},r_{\beta}\right)+\sum_{\alpha=1}^{N_{d}}E^D_{self}\left(\mathbf{b}_{\alpha},r_{\alpha}\right)+\sum_{\alpha=1}^{N_{d}}E^D_{relax}\left(\mathbf{b}_{\alpha},r_{\alpha}\right) ,
\label{total}
\end{eqnarray}
where $E_0$ is the energy of the defect-free, axisymmetric state,  
\begin{eqnarray}
E_{0}&=&\pi\int_{0}^{W}\sigma^{0}_{ij}u^{0}_{ij}-2\pi W T  u_{r}(W)\nonumber\\
&=& \frac{\pi W^2}{Y}\left((\nu-1)T^2+T Y \frac{W^2}{4R^2}+\frac{1}{384}\frac{Y^2 W^4}{R^4}\right) .
\end{eqnarray}

%This part of the elastic energy minus the work of adhesion (second term in eq .(1)) is, $E_{axi}\approx\pi W^2 T^2/Y\left(1/6+(64/3)\delta T/T\right)$.   

\subsection{Energy minimization of multi-dislocation ground states}
Here, we detail the numerical approach for exploring the multi-dislocation ground states. For a given $W/b$ ratio and curvature $K_{G}W^2$ (which correspond to a given value of $\epsilon$),  total dislocation number $N_{d}$ and reduced tension $T/T^{*}$, two classes of simulations were performed:  1) ``free dislocation" and 2) ``$n$-fold" simulations. ``Free dislocation" simulations start with random initial configurations of $N_{d}$ dislocation coordinates at $(r_{i}, \phi_{i})$, for $i=1...N_{d}$, with $\mathbf{b}=b\hat{\theta}$. Each simulation starts with $10^3-10^4$ random initial configurations (depending on the number of dislocations growing with $\epsilon^{-1/2}$) to account for the large number of local minima for $N_d \gg1$. The total  energy in eq. (\ref{total}) is minimized with respect to the position of the dislocations for all randomly initiated copies using the method of steepest descent.  The state with lowest resulting energy for a given $N_d$ is selected as the minimal energy for $N_d$ dislocation, $E(N_d; T/T_*, b/W, K_G W^2)$. In order to find the minimal-energy dislocation number, we determine $E(N_d; T/T_*, b/W, K_G W^2)$ for a range of possible dislocation numbers, $N_{d}=N^{c}_{d}\pm 0.25 N^{c}_{d}$ (where we use a linearized approximation of the continuum theory prediction $N^{c}\equiv\epsilon^{-1/2}\left(1-T/T_*\right)$ as the initial guess) and select the $N_d$ corresponding to lowest energy.  The resulting ``simulated ground states" are structures that are minimized with respect to the dislocation positions and dislocation number. 

For the case of fixed ``$n$-fold" simulations,  dislocations are constrained to $n_{s}$ identical radial lines (scars),  equally spaced at angular intervals of $2\pi/n_{s}$ on the cap. The radial positions of the $N_d/n_s=M$ concentric rings (constrained to an integer) of dislocations are initialized randomly, then relaxed via steepest descent.  Similar to the procedure outlined for ``free dislocation" simulations, the scar number is varied to find the optimal $n_{s}$ for a given $N_d$, $T/T_*$,  $b/W$ and $K_G W^2$. Both ``$n$-fold" and ``free dislocation" simulations are performed in the range of $T/T_{*}=0...1$, with a step size $\delta T/T_{*}=0.05$, for $\epsilon=0.17\times 10^{-4}-0.15\times10^{-2}$. These simulations were carried out over a range of cap sizes and curvatures: $W/b=100-1400$ and $W/R=0.05-0.3$ (see Table \ref{sims} for full list parameter values).

\begin{table}[ht] 
\caption{Parameters used for ground-state simulations} % title of Table 
\label{sims}
\centering % used for centering table 
\begin{tabular}{|c | c | c | c |} % centered columns (4 columns) 
\hline\hline %inserts double horizontal lines 
$b/W$  & $W/R$ & $\epsilon_{d}^{-1/2}$ & Simulation type \\ [0.5ex] % inserts table 
%heading 
\hline % inserts single horizontal line 

0.005  & 0.25 & 50 & Free \\ 
\hline
0.0011 & 0.07 & 63 & Free \\ 
\hline
0.0011 & 0.09 & 81 & Free \\ 
\hline
0.0011 & 0.11 & 99 &  Free \\ \hline
0.0011 & 0.13 & 117 & Free \\ \hline
0.001 & 0.15 & 150 & Free \\ \hline
0.0011 & 0.19 & 171 & Free \\ \hline
0.0011 & 0.21 & 189 & Free \\ \hline
0.0011 & 0.23 & 207 & Free \\ \hline
0.0011 & 0.25 & 225 & Free \\ \hline
0.01  & 0.25 & 25 & $n$-fold \\ \hline
0.01  & 0.3 & 30 & $n$-fold \\ \hline

0.005  & 0.25 & 50 & $n$-fold  \\ \hline
0.0011 & 0.07 & 63 & $n$-fold  \\ \hline
0.0011 & 0.09 & 81 & $n$-fold \\ \hline
0.0011 & 0.11 & 99 &  $n$-fold  \\ \hline
0.0011 & 0.13 & 117 & $n$-fold  \\ \hline
0.001 & 0.15 & 150 & $n$-fold  \\ \hline
0.0011 & 0.19 & 171 & $n$-fold  \\ \hline
0.0011 & 0.21 & 189 & $n$-fold  \\ \hline
0.0011 & 0.23 & 207 & $n$-fold  \\ \hline
0.0011 & 0.25 & 225 & $n$-fold  \\ \hline
0.0007  & 0.05  &  70  & $n$-fold \\ \hline
0.0007  & 0.06  &  84  & $n$-fold \\ \hline
0.0007  & 0.07 & 98 & $n$-fold \\ \hline
0.0007  & 0.08 & 112 & $n$-fold \\ \hline
0.0007  & 0.09 & 126 & $n$-fold \\ \hline
0.0007  & 0.1 & 140 & $n$-fold \\ \hline
0.0007  & 0.11 & 154 & $n$-fold \\ \hline

0.0007  & 0.12 & 154 & $n$-fold \\ \hline
0.0007  & 0.13 & 182 & $n$-fold \\ \hline
0.0007  & 0.14 & 196 & $n$-fold \\ \hline
0.0007  & 0.15 & 210 & $n$-fold \\ \hline
0.0007  & 0.17 & 238 & $n$-fold \\ \hline
0.0007  & 0.19 & 266  & $n$-fold \\   [1ex] % [1ex] adds vertical space 
\hline %inserts single line 
\end{tabular} 
\label{table:nonlin} % is used to refer this table in the text 
\end{table}

\subsection{Self-energy of scars}
In \cite{grason_davidovitch} it was shown for the weak confinement regime ($T\rightarrow T_{*}$), and argued more generally in the main text, that the subdominant energetics associated with the {\it self-energies of scars}  is responsible for selecting the optimal symmetry of $n$-fold scar patterns.  In the main text a scaling prediction for the $n_s$ dependence was made based on the distinct energetics associated with scar lengths and scar ends.  Here, we derive an explicit expression for the self energy contribution of scars in terms of dislocation energetics (self-energies and interactions) which we then minimize numerically with respect to $n_s$ to find a prediction for optimal scar number, $n_{s}\left(T/T_*\right)$, for arbitrary value of $T/T_*$.  Because the dominant pattern of stress $\sigma_{ij}^d$ and continuum limit defect-distribution $\rho(r)$ are independent of $n_s$, scar number enters the self-energy calculation of scars only through the change in linear density of dislocations along a scar, $\lambda(r)$.  Assuming pattern of $n_s$-fold symmetry we find a local dislocation spacing $D(r) = 1/\lambda(r)$,
\begin{equation}
D(r) = \frac{ n_s }{ 2 \pi r \rho(r)} ,
\label{Dr}
\end{equation}
which shows that scars become more diffuse (dense) lengthwise as their number increases (decreases).

The self-energy of a scar derives from the sum of the self-energies of individual dislocations and the sum over all pairwise interactions between dislocations along a single scar. For the case, of parallel dislocation pairs along a single scar, the form of dislocation interaction simplifies to,
\begin{equation}
E^D_{dis}(r_1,r_2)=\frac{Y b^{2}}{4 \pi^2 } \bigg[-\frac{1}{2}(\ln \cos ^{2}\xi+\sin^{2}\xi) - \sin^2 \xi  (1-r_1r_2/W^2) \bigg],
\end{equation}
where $\cos \xi = W|r_1-r_2|/(W^2-r_1r_2)$.  The total contribution from the self-energies of the $n_s$ scars can be written as the summations
\begin{equation}
E_{\rm self}/n_s= \sum_{\alpha=2}^{M}\sum_{\beta<\alpha}^{M} E^D_{int}\left(r_{\alpha},r_{\beta}\right)+\sum_{\alpha=1}^{M} E^D_{self}\left(r_{\alpha}\right) ,
\end{equation}
where $M=N_d/n_s$ is the number of dislocation per scar.  To approximate the value of the discrete sums along the scar, we replace dislocation self-energies and interaction energies with their {\it mean values} along intervals of width $D(r_\alpha)$, centered around dislocation positions $r_\alpha$, allowing us to convert sums to integrals,
\begin{eqnarray}
E_{\rm self}/n_s	    &\cong&\sum_{\alpha=2}^{M}\sum_{\beta<\alpha}^{M} \frac{1}{D(r_\beta)}\int_{r_{\beta}-D(r_\beta)/2}^{r_{\beta}+D(r_\beta)/2} E^D_{int}\left(r,r_{\alpha}\right)dr+\sum_{\alpha=1}^{M}  \frac{1}{D(r_\alpha)}\int_{r_{\alpha}-D(r_\alpha)/2}^{r_{\alpha}+D(r_\alpha)/2} E^D_{self}\left(r\right)dr
	\nonumber\\
             &\cong&\sum_{\alpha=2}^{M} \int_{r_\alpha+D(r_\alpha)/2}^{W+D(W)/2}\lambda(r) E^D_{dis}\left(r,r_{\alpha}\right)dr+\int_{L-D(L)/2}^{W+D(W)/2}\lambda(r) E^D_{self}\left(r\right)dr
	\nonumber\\
              &=&\int_{L}^{W}\lambda(r')dr' \int_{r'+D(r')/2}^{W}\lambda(r) E^D_{int}\left(r,r'\right)dr+\int_{L}^{W}\lambda(r,T/T_*) E^D_{self}\left(r\right)dr,
\end{eqnarray}
where we have dropped $\pm D/2$ corrections to the range of integration the ends of scars $r=L$ and $r=W$.  Substituting eq. (\ref{Dr}), and defining $\epsilon^{1/2} n_s d(r) = D(r)$ to scale out the $\epsilon$ and $n_s$ dependence of dislocation spacing, we find the total self-energy of scars as a function of scar number,
\begin{equation}
E_{\rm self} (n_s)= \frac{4 \pi^2 }{n_s} \int_L^W \rho(r') r' dr' \int_{r'+n_s \epsilon^{1/2} d(r') }^W \rho(r) E^D_{int} (r,r') r dr + 2 \pi \int_L^W \rho(r) E^D_{self}(r) r dr .
\label{Ens}
\end{equation}
Since $\rho(r)$ and $L$ are independent of scar number, the second term, which represents the contribution from dislocation self-energies along the scar, is independent of $n_s$, while the $n_s$-dependence of the first term --- deriving from pairwise dislocation interactions --- derives from the numerator as well as $n_s$-dependence limit of integration over $r$.   

Careful inspection of eq. (\ref{Ens}) shows  it to be a function (up to a multiplicative constant) of two dimensionless variables, reduced tension $T/T_*$ and scaled scar number $\bar{n}_s = \epsilon^{1/2} n_s$.  To determine the optimal scar number, $E_{\rm self} (n_s)$, is numerically integrated, and numerically minimized with respect to $\bar{n}_s$ for a given $T/T_*$ to determine the function $\bar{n}_s(T/T_*)$ plotted in Fig. 2c of the main text.

\subsection{Dominant energy}
Here, we compute the form of the {\it dominant} energy stored in the elastic energy of the asymptotic stress pattern, $\sigma^{\rm d}_{ij}$, which is realized in the singular, continuum limit, in order to extract and compare the {\it subdominant} energetics of $n$-fold and ``forked scar" dislocation morphologies observed in our simulations.  The dominant energy follows from the solution of stress, strain and displacement fields corresponding to, $\sigma^{\rm d}_{ij}$, solutions which are split into two zones, defect free zone for $r< L$, $\sigma_{\theta \theta} >0$ and compression free zone $r\geq L$, $\sigma_{\theta \theta} =0$:
\begin{eqnarray}
E_{dom}=\pi\int_{0}^{L(T/T_{*})}(\sigma^{in}_{rr}u^{in}_{rr}+ \sigma^{in}_{\theta \theta }u^{in}_{\theta \theta })  r dr+\pi\int_{L(T/T^{*})}^{W}\sigma^{out}_{rr}u^{out}_{rr}r dr-2\pi W T  u_{r}(W),
%&=&
%\pi W^2T^2\int_{(T/T^*)^{1/3}W}^{W}\frac{dr}{r}-2\pi W T u_{r}(W) 
\end{eqnarray}
In outer zone we have the compression free solution for stress $\sigma^{out}_{rr}=TW/r$, and strain $u^{out}_{rr}=\frac{T}{Y}W/r$. On the other hand the geometric strain-displacement relation yields, 
\begin{eqnarray}
u_{rr}^{out}=\partial_{r}u^{out}_{r}+1/2(r/R)^2,
\label{rstrain}
\end{eqnarray}
where $R$ is the radius of the curvature. Knowing $u^{out}_{rr}$, we integrate eq. (\ref{rstrain}) yielding, $u^{out}_{r}=TW/Y \ln (r/C_0)-1/6(r^{3}/R^2)$. $C_0$ is determined by matching $u_{r}$ at edge of the scarred (compression free) zone, $r=L(T/T_*)=(T/T_*)^{1/3}W$. To find $u_{r}^{in}$ we start with the stress solutions for the inner zone of the sheet that is the rescaled version of stress distribution of the axisymmetric state (eq. (3) in the manuscript), with $W\rightarrow L$ and $T^{*}\rightarrow T(T/T_{*})^{-1/3}$. Integration of the radial strain subject to $u_r(0) =0$ yeilds,
\begin{eqnarray}
u_{r}^{in}=-\frac{T}{Y}(\nu-1)r+ \frac{(\nu-3)r^3}{16R^2}-\frac{(\nu-1)W^2r}{16R^2},
\end{eqnarray}
where $\nu$ is poisson ratio.

From matching condition $u_{r}^{in}(L)=u_{r}^{out}(L)$, we find the constant,  $C_{0}=2\exp\left(\nu-4/3\right)R^{2/3}\left(\frac{TW}{Y}\right)^{1/3}$. Now we can calculate $E_{dom}$ in eq. (13) to find

\begin{eqnarray}
E_{dom}=\frac{\pi T W^2}{6Y} \left(3(2\nu-3)T+\frac{2W^2}{R^2} Y+2 T\ln\left[2\frac{R^2 T}{W^2 Y}\right]\right),
\end{eqnarray}
We  calculate the sub-dominant energy of ``free dislocation" and $n-$fold simulations simply by subtracting the dominant energy (eq. (17)) from total energy of the system in eq. (8)
\begin{eqnarray}
\label{Esub}
E_{sub}&=&E_{tot}-E_{dom}\nonumber\\
&=&\sum_{\alpha=2}^{N_{d}}\sum_{\beta<\alpha}^{N_{d}} E^D_{int}\left(\mathbf{b}_{\alpha},r_{\alpha};\mathbf{b}_{\beta},r_{\beta}\right)+\sum_{\alpha=1}^{N_{d}}E^D_{self}\left(\mathbf{b}_{\alpha},r_{\alpha}\right)+\sum_{\alpha=1}^{N_{d}}E^D_{relax}\left(\mathbf{b}_{\alpha},r_{\alpha}\right)-(E_{dom}-E_{0}) .
\end{eqnarray}
The last term in eq. (\ref{Esub}) is,
\begin{eqnarray}
E_{dom}-E_{0}=\frac{\pi W^{2}}{6Y} T_*^{2}\left[-1+(4-3\alpha)\alpha+2\alpha^{2}\ln\alpha\right],
\end{eqnarray}
where we define $\alpha \equiv T/T_*$.  Expanding the above expression in the limit of weak confinement, $T\rightarrow T_{*}$ we have $E_{dom}-E_{0}\approx \frac{\pi W^2T_*(\alpha - 1) ^3}{9Y}$.
The first three sums in eq. (19) are calculated explicitly in numerical simulations of caps.

\begin{figure}

\includegraphics[width=0.875\textwidth]{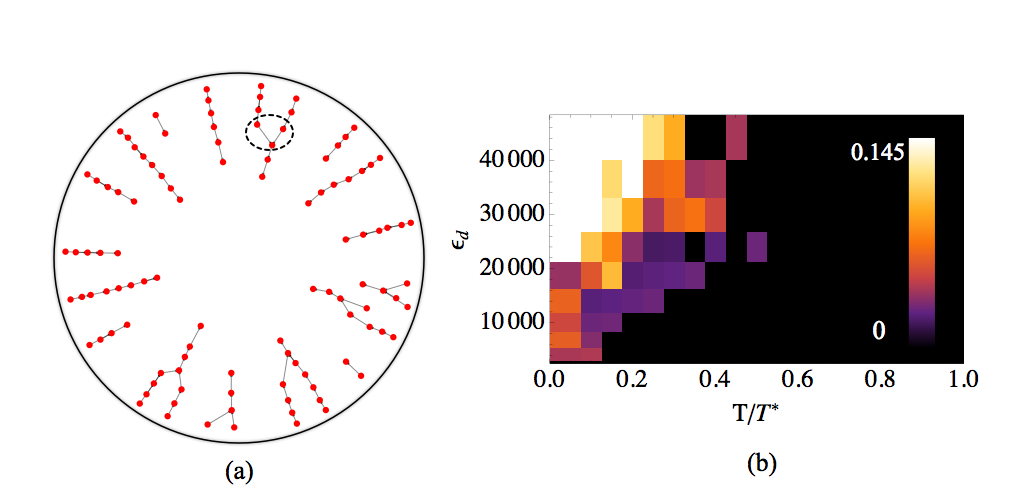}
\caption{(a) shows the cluster analysis which groups dislocations into scars, and counts the number of ``forks" or branches (highlighted in the dashed circle).  (b) shows the map of $\rho_{F}$, the number of forks per scar found in multi-dislocation simulations.}
\label{forked}
\end{figure}  

\subsection{Structural analysis of polydisperse, forked-scars}
Here we show that the boundary between symmetric, $n$-fold scars at small $T$ and nonsymmetric structures at larger $T$  observed in Fig. 3 of the main text does not depend significantly on our chosen structural measure of $n$-fold symmetry, $S$ the ratio between the first two peaks in the angular Fourier spectrum. Alternatively, we can quantify the transition in terms of the number of ``forks" or branches appearing in each optimal configuration. To count the number of forks, we use a simple clustering algorithm that counts number of scars by recognizing set of neighboring dislocations as a an individual scar according to the following rules. 1) each dislocation finds just one nearest neighboring dislocation at a smaller radius, within a $\delta\phi=\pi/4$ azimuthal interval from the radial direction, or 5-7 dipole.  2) each cluster (scar) is a group of dislocations which share at least one neighbor. 3) We define a ``fork" as a dislocation that is the neighbor of two or more dislocations at larger radii. One example of a fork is shown in Fig.\ref{forked}a in the dashed circle where filled circles represent dislocations in an optimal configuration. Hence dislocation clusters with perfect radial arrays of dislocations have no forks.  We find that when a $n$-fold symmetry radial scars become sufficiently unstable, scars become increasing branched.  We quantify the degree of ``scar heterogeneity" in terms of number of forks per dislocation, $\rho_{F}=F/N_{d}$. In Fig. ref{forked} we show the map of the scar fork density $\rho_{F}$, for free-dislocation, ground-state configurations. In this phase map black color shows regions of zero fork density and lighter colors show regions of branched polydisperse scarred patterns, notably a highlighting nearly identical  regions of ordered/disordered scars as shown by the Fourier analysis of dislocation distribution.


\begin{thebibliography}{100}




%\bibitem{aste_weaire}
%T. Aste and D. L. Weaire, {\it The Pursuit of Perfect Packing}, 2nd ed. (Taylor and Francis, New York,  2008).



\bibitem{caspar}
D. L. D. Caspar and A. Klug, Cold Spring Harbor Symp. Quant. Biol. {\bf 27}, 1 (1962).

\bibitem{bruinsma_pnas}
R. Zandi, D. Reguara, R. F. Bruinsma, W. M. Gelbart and J. Rudnick, Proc. Nat. Acad. Sci. USA {\bf 101}, 15556 (2004).

\bibitem{gompper}
S. Schneider and G. Gompper, Europhys. Lett. {\bf 70}, 136 (2005).

\bibitem{olvera}
G. Vernizzi, R. Sknepnek and M. Olvera de la Cruz, Proc. Nat. Acad. Sci. USA {\bf 108}, 4292 (2010).

\bibitem{bausch}
A. R. Bausch, M. J. Bowick, A. Cacciuto, A. D. Dinsmore, M. F. Hsu, D. R. Nelson, M. G. Nikolaides, A. Travesset and D. A. Weitz, Science {\bf 299}, 1716 (2003).

\bibitem{kleman}
M. Kl\'eman, Adv. Phys. {\bf 38}, 605 (1989).

\bibitem{sadoc}
J.-F. Sadoc and R. Mosseri, {\it Geometrical Frustration} (Cambridge
University Press, Cambridge, 1999).

\bibitem{altschuler}
E. L. Altschuler, T. J. Williams, E. R. Ratner, F. Dowla and F. Wooten, Phys. Rev. Lett. {\bf 72}, 2671 (1994).


\bibitem{bowick_caccuito}
M. Bowick, A. Caccuito, D. R. Nelson and A. Travesset, Phys. Rev. Lett. {\bf 89}, 185502 (2002).



\bibitem{irvine}
W. T. Irvine, V. Vitelli and P. M. Chaikin, Nature {\bf 468}, 947 (2011).



\bibitem{wales_09}
D. J. Wales, H. McKay and E. L. Altschuler, Phys. Rev. B {\bf 79}, 224115 (2009).


\bibitem{wales_13}
H. Kusumaatmaja and D. J. Wales, Phys. Rev. Lett. {\bf 110}, 165502 (2013).



\bibitem{bowick}
M. Bowick, D. R. Nelson and A. Travesset, Phys. Rev. B {\bf 62}, 8738 (2000).

\bibitem{travesset}
A. Travesset, Phys. Rev. B {\bf 68}, 115421 (2003).

\bibitem{vitelli_lucks}
V. Vitelli, J. B. Lucks and D. R. Nelson, Proc. Nat. Acad. Sci USA {\bf 103}, 12323 (2006).


\bibitem{grason_davidovitch}
G. M. Grason and B. Davidovitch, Proc. Nat. Acad. Sci. USA {\bf 110}, 12893 (2013).

\bibitem{cerda_maha}
E. Cerda and L. Mahadevan, Phys. Rev. Lett. {\bf 90}, 074302 (2003).

\bibitem{king}
H. King, R. D. Schroll, B. Davidovitch, Proc. Nat. Acad. Sci. USA {\bf 109}, 9716 (2012).

\bibitem{nelson_book}
D. R. Nelson, {\it Defects and Geometry in Condensed Matter Physics}, (Cambridge, Cambridge, 2002).


\bibitem{grason_prl_10}
G. M. Grason, Phy. Rev. Lett. {\bf 105}, 045502 (2010).

\bibitem{grason_12}
G. M. Grason, Phy. Rev. E {\bf 85}, 031603 (2012).

\bibitem{azadi_grason_12}
A. Azadi and G. M. Grason, Phy. Rev. E {\bf 85}, 031604 (2012).

\bibitem{hirth}
J. P. Hirth and J. Lothe, {\it Theory of Dislocations}, (Wiley, New York, 1982).

\bibitem{schroll}
B. Davidovitch, R. D. Schroll, D. Vella, M. Adda-Bedia and E. A. Cerda, {\bf 108}, 18227 (2011).

\bibitem{azadi_tobe}
A. Azadi and G. M. Grason, to be published.

%\bibitem{nelson_seung}
%H. S. Seung and D. R. Nelson, Phys. Rev. A {\bf 38}, 1005 (1988).


%\bibitem{nelson_peliti}
%D. R. Nelson and L. Peliti, J. Phys. France {\bf 48}, 1085 (1987).

\end{thebibliography}
\end{document}